\documentstyle[preprint,aps]{revtex}
\begin{document}
\draft
\title{\Large Pseudoclassical supersymmetrical model for 2+1 Dirac particle}
\author{D.M. Gitman and A.E. Gon\c calves }
\address{Instituto de F\'{\i}sica, Universidade de S\~ao Paulo\\
P.O. Box 66318, 05389-970 S\~ao Paulo, SP, Brazil}
\author{ I.V. Tyutin}
\address{Lebedev Physical Institute, 117924 Moscow, Russia}

\date{\today}
\maketitle

\begin{abstract}
A new pseudoclassical supersymmetrical model of a spinning 
particle in $2+1$ dimensions is proposed. Different ways of its 
quantization are discussed. They all reproduce the minimal quantum 
theory of the particle.
\end{abstract}

In this paper we present a new pseudoclassical model for a massive Dirac
particle in $2+1$ dimensions. Besides a pure theoretical interest to complete the
theory of relativistic particles,  there is a
direct relation  with the $2+1$ field theory \cite{JTD}, which attracts in recent
years great attention due to various reasons: e.g. because of nontrivial
topological properties, and especially due to a possibility of the
existence of particles with fractional spins and exotic statistics
(anyons), having probably applications to fractional Hall effect,
high-$T_c$ superconductivity and so on \cite{Wil}. The well known 
pseudoclassical supersymmetrical model for
Dirac (spinning) particle in $3+1$ dimensions was proposed and
investigated by several authors  \cite{BM}. Attempts to extend the pseudoclassical 
description of spinning particle to the arbitrary odd-dimensions case
had met some problems, which are connected with the absence of an analog of
$\gamma^5$-matrix in odd-dimensions. For instance, in $2+1$ dimensions the
direct generalization of the Berezin-Marinov action (standart action) 
does not reproduce a
minimal quantum theory of spinning particle, which has to provide only one
value of the spin projection ($1/2$ or $-1/2$). In papers \cite{P} they
have proposed
two modifications of the standard action to get such a minimal theory, but
they can not be considered as satisfactory solutions of the problem.
The first action \cite{P} is classically equivalent to the standard action
and does not provide required quantum properties in course of canonical and
path-integral quantization.  Moreover, it is  $P$- and $T$-invariant, so
that an anomaly is present. Another one \cite{P} does not obey gauge
supersymmetries and therefore loses the main attractive features in such kind
of models, which allows one to treat them as prototypes of superstrings or
some modes in superstring theory.
The action, we are proposing, is invriant under three gauge transformations: 
reparametrization and two supertransformations. It is $P$- and
$T$-noninvariant in full accordance with the expected properties of the 
minimal theory in $2+1$ dimensions. Dirac
quantization and quasi-canonical quantization with fixation of the gauge freedom, which
corresponds to two types of gauge transformations of the 
three existing, both lead to
the quantum theory of spinning particle in $2+1$ dimensions. 

The new action to describe a Dirac particle in $2+1$
dimensions has the form 
\begin{eqnarray}\label{act}
\displaystyle
S &=&\int\limits_0^1\left[-\frac{z^2}{2e}-e\frac{m^2}{2}-g\dot x^\mu A_\mu+
\imath geF_{\mu\nu}\psi^\mu\psi^\nu-
\imath m\psi^3\chi-\frac{1}{2}sm\kappa-
\imath\psi_a\dot\psi^a\right]d\tau \nonumber\\ 
\displaystyle
&\equiv &\int\limits_0^1Ld\tau\;,\;\; s=\pm\;;\;\;
z^\mu=\dot x^\mu-\imath\psi^\mu\chi+\imath
\varepsilon^{\mu\nu\lambda}\psi_\nu\psi_\lambda\kappa\;;
\end{eqnarray}
the Latin indices $a$, $b$, $c$, $\ldots$, run over 0, 1, 2, 3, whereas the
Greek (Lorentz) ones $\mu$, $\nu$, $\ldots$, run over 0, 1, 2; $x^\mu$, $e$,
$\kappa$ are even and $\psi^a$, $\chi$ are odd variables; 
$F_{\mu\nu}=\partial_\mu
A_\nu-\partial_\nu A_\mu$ is the strength tensor, $g$ is the
$U(1)$-charge of the particle, interacting with an external gauge field
$A_\mu(x)$, which can have the Maxwell or (and) Chern-Simons nature;
$\varepsilon^{\mu\nu\lambda}$ is the totally antisymmetric tensor density of
Levi-Civita in $2+1$ dimensions normalized by $\varepsilon^{012}=+1$;
$\eta_{ab}=\hbox{diag}(1,-1,-1,-1)$, $\eta_{\mu\nu}=\hbox{diag}(1,-1,-1)$. We
suppose that $x^\mu$ and $\psi^\mu$ are $2+1$ Lorentz vectors and $e$,
$\kappa$, $\psi^3$, $\chi$ are scalars, so that the action (\ref{act}) 
is invariant under the restricted Lorentz transformations 
(but not $P$- and $T$-invariant). It is invariant
under reparametrizations and two types of gauge supertransformations: 
\begin{eqnarray}\label{rep}
&&\delta x^\mu=\dot x^\mu\xi\;,\;\; \delta e=\frac{d}{d\tau}(e\xi)\;,
\;\;\delta\psi^a=\dot\psi^a\xi\;, \;\;
\delta\chi=\frac{d}{d\tau}(\chi\xi)\;, \;\;
\delta\kappa=\frac{d}{d\tau}(\kappa\xi)\;, \\
&&\delta x^\mu=\imath\psi^\mu \epsilon\;,\;\;\delta e=\imath\chi\epsilon\;,\;\;
\delta\psi^\mu=\frac{z^\mu}{2e}\epsilon\;,\;\;
\delta\psi^3=\frac{m}{2}\epsilon,\;\;\delta\chi=\dot\epsilon\;,
\;\;\delta\kappa=0\;, \label{st1}\\ 
&&\delta x^\mu=-\imath\varepsilon^{\mu\nu\lambda}\psi_\nu\psi_\lambda\theta
\;,\;\;
\delta\psi^\mu=\frac{1}{e}\varepsilon^{\mu\nu\lambda}z_\nu\psi_\lambda\theta\;,
\;\;\delta\kappa=\dot\theta,\;\; \delta e=\delta\psi^3=\delta\chi=0\;,
\label{st2}
\end{eqnarray}
where $\xi(\tau),\;\theta(\tau)$ are even, and $\epsilon(\tau)$ is
odd parameter.

Going over to the Hamiltonian formulation, we introduce the canonical momenta
\begin{eqnarray}\label{cm}
\displaystyle
&&\pi_\mu=\frac{\partial L}{\partial\dot x^\mu}=-\frac{1}{e}z_\mu-gA_\mu\;,\;\;
P_e=\frac{\partial L}{\partial\dot e}=0\;, \;\;
P_\chi=\frac{\partial_rL}{\partial\dot\chi}=0\;, \nonumber\\
\displaystyle
&&P_\kappa=\frac{\partial L}{\partial\dot\kappa}=0\;,\;\;
P_a=\frac{\partial_rL}{\partial\dot\psi^a}=-\imath\psi_a\;.
\end{eqnarray}
It follows from (\ref{cm}) that there exist primary constraints
$\Phi^{(1)}=0 \;(\Phi^{(1)}_1=P_e\;, \;\; \Phi^{(1)}_2=P_\chi\;, \;\;
\Phi^{(1)}_3=P_\kappa\;,\;\; \Phi^{(1)}_{4a}=P_a+\imath\psi_a$).   
Constructing the total Hamiltonian $H^{(1)}$, according to the standard
procedure \cite{D,GT1}, we get $H^{(1)}=H+\lambda_A\Phi^{(1)}_A$ with  
\begin{equation}\label{ham}
H=-\frac{e}{2}(\Pi^2-m^2+2\imath gF_{\mu\nu}\psi^\mu\psi^\nu)+
\imath(\Pi_\mu\psi^\mu+m\psi^3)\chi -\imath
(\varepsilon^{\mu\nu\lambda}\Pi_\mu\psi_\nu\psi_\lambda +
\frac{\imath}{2}sm)\kappa\;,
\end{equation}
where $\Pi_\mu=\pi_\mu+gA_\mu$.
From the consistency conditions $\dot\Phi^{(1)}=\{\Phi^{(1)},H^{(1)}\}=0$ 
we find secondary constraints $\Phi^{(2)}=0 \;\;
(\Phi^{(2)}_1=\Pi_\mu\psi^\mu+m\psi^3\;,\;\; \Phi^{(2)}_2 =
\Pi^2-m^2+2\imath gF_{\mu\nu}\psi^\mu\psi^\nu,\;\;
\Phi^{(2)}_3=\varepsilon^{\mu\nu\lambda}\Pi_\mu\psi_\nu\psi_\lambda+
\frac{\imath}{2}sm$),  and determine $\lambda$, which correspond to 
the primary constraints
$\Phi^{(1)}_{4}$. No more secondary constraints arise from the
consistency conditions and the Lagrangian multipliers, correspondent to the
primary constraints $\Phi^{(1)}_i$, $i=1,2,3$, remain undetermined. The
Hamiltonian (\ref{ham}) is proportional to the constraints.  
One can go over from the initial set of constraints $\Phi^{(1)},
\Phi^{(2)}$ to the equivalent ones $\Phi^{(1)},\tilde{\Phi}^{(2)}$,
where $\tilde{\Phi}{}^{(2)}=\Phi^{(2)}\left(\psi\rightarrow\tilde{\psi}=\psi+
\frac{\imath}{2}\Phi^{(1)}_4 \right)$.  
The new set of constraints can be explicitly divided in a set of the
first-class constraints, which are ($\Phi^{(1)}_i$, $i=1,2,3$,
$\tilde{\Phi}^{(2)}$) and in a set of second-class constraints
$\Phi^{(1)}_4$. 

Let us consider the Dirac quantization, where the second-class
constraints define the Dirac brackets and therefore the commutation relations,
whereas, the first-class constraints, being applied to the state vectors,
define physical states. For essential operators and nonzeroth
commutation relations one can obtain in the case of consideration: 
\begin{equation}\label{cr}
[\hat{x}^\mu,\hat{\pi}_\nu]=\imath\{x^\mu,\pi_\nu\}_{D(\Phi^{(1)}_4)}=
\imath\delta^\mu_\nu\;, \;\;
[\hat{\psi}^a,\hat{\psi}^b]_+=\imath\{\psi^a,\psi^b\}_{D(\Phi^{(1)}_4)}=
-\frac{1}{2}\eta^{ab}\;. 
\end{equation}
It is possible to construct a realization of the commutation relations
(\ref{cr}) in a Hilbert space ${\cal R}$ whose elements ${\bf f}\in 
{\cal R}$ are 
four-component columns dependent on $x$,
\begin{equation}\label{fcc}
{\bf f}(x)=\left(\begin{array}{c}u_-(x)\\u_+(x)\end{array}\right),\;\;
\hat{x}^\mu=x^\mu{\bf I}\;,\;\; \hat{\pi}_\mu=-\imath\partial_\mu{\bf I}\;,
\;\; \hat{\psi}^a=\frac{\imath}{2}\gamma^a, 
\end{equation}
where $u_\mp(x)$ are two-component columns, and ${\bf I}$ is
$4\times 4$ unit matrix; $\gamma^a$, $a=0,1,2,3,$ are $\gamma$-matrices
in $3+1$ dimensions, which we select in the spinor representation
$\gamma^0={\rm antidiag}(I,\;I),\;\;\gamma^i={\rm antidiag}
(\sigma^i,\;-\sigma^i),\;\; i=1,2,3\;,\;\; 
\sigma^i$ are the Pauli matrices, and $I$ is $2\times 2$ unit matrix.
According to the scheme of quantization selected, the operators of the 
first-class constraints have to be applied to the state vectors to
define physical sector, namely, $\hat{\Phi}^{(2)}{\bf f}(x)=0\;$, 
where $\hat{\Phi}^{(2)}$ are operators, which correspond to the constraints
$\Phi^{(2)}$. Taken into account (\ref{fcc}), one can write the equation  
$\hat{\Phi}^{(2)}_1{\bf f}(x)=0$ as 
\begin{equation}\label{setc}
\displaystyle
[(\imath\partial_\mu-gA_\mu)\gamma^\mu-m\gamma^3]{\bf f}(x)=0 \;\;
\Longleftrightarrow\left\{ 
\begin{array}{c}
[(\imath\partial_\mu-gA_\mu)\Gamma^\mu_+-m]u_+(x)=0\;, \nonumber\\
\displaystyle
[(\imath\partial_\mu-gA_\mu)\Gamma^\mu_-+m]u_-(x)=0\;,
\end{array}\right.
\end{equation}
where two sets of $\gamma$-matrices in 2+1 
dimensions are introduced, 
$\Gamma^0_s=\sigma^3, \;\; \Gamma^1_s=s\imath\sigma^2, \;\;
\Gamma^2_s=-s\imath\sigma^1\;,\;\; \;s=\pm\;\;
\displaystyle
\Gamma^\mu_-=\Gamma_{+\mu}\;, \;\;
[\Gamma^\mu_s,\Gamma^\nu_s]_+=2\eta^{\mu\nu}\;.$ 
Constructing the operator $\hat{\Phi}^{(2)}_2$ according to the
classical function $\Phi^{(2)}_2$, we meet an ordering
problem since the latter contains terms  of the form
$\pi_\mu A^\mu(x)$. For such terms we choose the symmetrized (Weyl) form of the
correspondent operators,  which provides, in particular, the consistency of 
two equations $\hat{\Phi}^{(2)}_1{\bf f}=0$ and $\hat{\Phi}^{(2)}_2{\bf f}=0$,
because of in this case $\hat{\Phi}^{(2)}_2=
(\hat{\Phi}^{(2)}_1)^2$.
The equation $\hat{\Phi}^{(2)}_3{\bf f}(x)=0$ can be presented in the 
following form 
\begin{equation}\label{ophi3}
\displaystyle
\left[\frac{1}{2}\varepsilon^{\mu\nu\lambda}(\imath\partial_\mu-gA_\mu)
\gamma_\nu\gamma_\lambda+\imath sm\right]{\bf f}(x)=0 \;\;
\Longleftrightarrow \;\;
\left\{
\begin{array}{c}\displaystyle
[(\imath\partial_\mu-gA_\mu)\Gamma^\mu_+-sm]u_+(x)=0\;, \\[9pt] \displaystyle
[(\imath\partial_\mu-gA_\mu)\Gamma^\mu_--sm]u_-(x)=0\;.
\end{array}
\right. 
\end{equation}
Combining eq. (\ref{setc}) and (\ref{ophi3}), we get 
\begin{equation}
[(\imath\partial_\mu-gA_\mu)\Gamma^\mu_s-sm]u_s(x)=0\;,\;\;
u_{-s}(x)\equiv 0\;,\;\;s=\pm\;.
\end{equation}

To interpret the quantum mechanics constructed one has to take into acount  
the operator, which corresponds to the angular momentum tensor 
$M_{\mu\nu}=x_\mu\pi_\nu-x_\nu\pi_\mu +i[\psi_\mu ,\psi_\nu].\;\;
\hat{M}_{\mu\nu}=-\imath(x_\mu\partial_\nu-x_\nu\partial_\mu)-\frac{\imath}{4}
\hbox{diag}\left([\Gamma_{-\mu},\Gamma_{-\nu}],\;
[\Gamma_{+\mu},\Gamma_{+\nu}]\right)$.
Thus one can see that the states with $(s=+)$ are
described by the two component wave function $u_+(x)$, which obeys the
Dirac equation in 2+1 dimensions and is transformed under the Lorentz 
transformation as spin $+1/2$ \cite{JN}.
For $(s=-)$ the quantization leads to the theory of 2+1 Dirac particle 
with spin $-1/2$ and wave function $u_-(x)$.

To quantize the theory canonically we have to impose as much as possible
supplementary gauge conditions to the first-class constraints. 
In the case under consideration, it turns out to be possible to 
impose gauge conditions to all the first-class constraints, excluding 
the constraint $\tilde{\Phi}^{(2)}_3$. Thus, we are fixing the gauge 
freedom, which  corresponds to two types of gauge transformations 
(\ref{rep}) and (\ref{st1}).  As a result  we remain only with one 
first-class constraint, which is the reduction of
$\Phi^{(2)}_3$ to the rest of constraints and gauge conditions. It can be
used to specify the physical states. All the second-class constraints 
form the Dirac brackets.  We consider below, for simplicity, 
the case without an
external field. The following gauge conditions $\Phi^G=0$ are imposed: 
$\Phi^G_1=e+\zeta\pi^{-1}_0\;,\;\; \Phi^G_2=\chi\;,\;\; \Phi^G_3=\kappa\;,\;\;
\Phi^G_4=x_0-\zeta\tau\;,\;\; \Phi^G_5=\psi^0\;, 
$ 
where $\zeta=-\hbox{sign}\,\pi^0$. (The gauge $x_0-\zeta\tau=0$ was
first proposed in \cite{GT2,GT1} as a conjugated gauge condition to
the constraint $\pi^2-m^2=0$). Using the consistency condition
$\dot\Phi^G=0$, one can determine the Lagrangian multipliers,
which correspond to the primary constraints $\Phi^{(1)}_i$,
$i=1,2,3$. 
To go over to a time-independent set of constraints (to use standart
scheme of quantization without modifications \cite{GT1}),  
we introduce the varialble $x^\prime_0,\;x^\prime_0=x_0-\zeta\tau$, 
instead of $x_0$, without changing the rest of the variables. 
That is a canonical transformation \cite{GT2}. 
The transformed Hamiltonian $H^{(1)\prime}$ is of the form
$H^{(1)\prime}=\omega+\{\Phi\}\;,\;\;
\omega=\sqrt{\pi^2_d+m^2}\;, \;\;\; d=1,2$, 
where $\{\Phi\}$ are terms proportional to the constraints and
$\omega$ is the physical Hamiltonian.
Now, all the constraints of the theory can be presented in the following 
equivalent form:  
$K=0$, $\phi=0$, $T=0$, where 
$K=(e-\omega^{-1}\;,\; P_e\;;\;\; \chi\;,\; P_\chi\;; \;\; \kappa\;,\; 
P_\kappa\;;\;\;
x^\prime_0\;,\; |\pi_0|-\omega\;; \;\; \psi^0\;,\; P_0),\;\; 
\phi=(\pi_d\psi^d+m\psi^3, P_k+\imath\psi_k)\;, \;\; d=1,2\;, \;\;
k=1,2,3$, and 
\begin{equation}\label{cc}
T=\zeta\omega[\psi_2,\psi_1]+\frac{\imath}{2}sm\;.
\end{equation}
The constraints $K$ and $\phi$ are of the second-class, whereas $T$ is the
first-class constraint.  Besides, the set $K$ has the so called special form
\cite{GT1}. In this case, if we eliminate the variables $e$, $P_e$, $\chi$,
$P_\chi$, $\kappa$, $P_\kappa$, $x^\prime_0$, $|\pi_0|$, $\psi^0$, and $P_0$,
using the constraints $K=0$, the Dirac brackets with respect to all
the second-class constraints $(K,\phi)$ reduce to ones
with respect to the constraints $\phi$ only. Thus, we can  
only consider the variables $x^d$, $\pi_d$, $\zeta$, $\psi^k$, $P_k$ and two
sets of constraints - the second-class ones $\phi$ and the first-class one
$T$. Nonzeroth Dirac brackets for the independent variables are
\begin{eqnarray}\label{db}
\displaystyle
&&\{x^d,\pi_r\}_{D(\phi)}=\delta^d_r\;, \;\;
\{x^d,x^r\}_{D(\phi)}=\frac{\imath}{\omega^2}[\psi^d,\psi^r]\;, \;\;
\{x^d,\psi^r\}_{D(\phi)}=-\frac{1}{\omega^2}\psi^d\pi_r\;, \nonumber\\ 
\displaystyle
&&\{\psi^d,\psi^r\}_{D(\phi)}=-\frac{\imath}{2}
(\delta^d_r-\omega^{-2}\pi_d\pi_r)\;,
\;\;d,r=1,2\;.
\end{eqnarray}
They define the commutation relations between the operators $\hat{x}^d$,
$\hat{\pi}_d$, $\hat{\psi}^d$, 
\begin{eqnarray}\label{rc}
\displaystyle
&&[\hat{x}^d,\hat{\pi}_r]=\imath\delta^d_r\;,\;\;
[\hat{x}^d,\hat{x}^r]=-\frac{1}{\hat{\omega}^2}[\hat{\psi}^d,\hat{\psi}^r]\;,
\nonumber\\
\displaystyle
&&[\hat{x}^d,\hat{\psi}^r]=-\frac{\imath}{\hat{\omega}^2}\hat{\psi}^d\hat{\pi}_r\;,
\;\;[\hat{\psi}^d,\hat{\psi}^r]_+=\frac{1}{2}(\delta^d_r-
\hat{\omega}^{-2}\hat{\pi}_d\hat{\pi}_r)\;.
\end{eqnarray}
We assume as usual \cite{GT2,GT1} the operator $\hat{\zeta}$ to have 
the eigenvalues
$\zeta=\pm1$ by analogy with the classical theory, so that $\hat{\zeta}^2=1$,
and also we assume the equations of the second-class constraints 
$\hat{\phi}=0$. Then one can
realize the algebra (\ref{rc}) in a Hilbert space ${\cal R}$, 
whose elements ${\bf f}\in {\cal R}$
are four-component columns dependent on ${\bf x}=(x^d)$, $d=1,2$,
\begin{eqnarray}\label{fcc1}
&&{\bf f}({\bf x})=\left(\begin{array}{c}f_+({\bf x})\\f_-({\bf x})
\end{array}\right)\;,\;\;
\hat{x}^d=x^d{\bf I}+\frac{1}{2\hat{\omega}^2}\varepsilon^{dr}
[\hat{\pi}_r\Sigma^3-m\Sigma^r]\;,\;\;\hat{\pi}_d=-\imath\partial_d{\bf I}\;,
\nonumber\\
&&\hat{\psi}^d=\frac{1}{2}(\delta^d_r-\hat{\omega}^{-2}\hat{\pi}_d\hat{\pi}_r)
\Sigma^r-\frac{1}{2\hat{\omega}^2}m\hat{\pi}_d\Sigma^3, \;\;
\hat{\zeta}=\hbox{antidiag}(I\;, -I)\;,
\end{eqnarray}
where $f_+({\bf x})$ and $f_-({\bf x)}$ are two-component columns, and 
$\Sigma^k=\hbox{diag}( \sigma^k,\sigma^k)$.
The operator $\hat{T}$ correspondent to the 
first-class constraint (\ref{cc}) specifies the physical states,  
\begin{equation}\label{T}
\hat{T}{\bf f}=0\;,\;\;
\hat{T}=\frac{\imath sm}{2\hat{\omega}}\hat{\zeta}\Sigma^3\left[
\hat{\zeta}\hat{\omega}\Sigma^3+\imath\partial_1(\imath s\Sigma^2)+
\imath\partial_2(-\imath s\Sigma^1)-sm\right]\;.
\end{equation}
Besides theses states  obey the Schr\"odinger
equation, which defines their ``time'' dependence, 
$(\imath\partial/\partial\tau-\hat{\omega}){\bf f}=0\;,\;\;
\hat{\omega}=\sqrt{\hat{\pi}^2_d+m^2}$, where the quantum
Hamiltonian $\hat{\omega}$ corresponds the classical one $\omega$.  
Introducing the
physical time $x^0=\zeta\tau$ instead of the parameter $\tau$ \cite{GT2,GT1},
we can rewrite the Schr\"odinger equation in the following form 
$(\imath{\partial/\partial x^0}-\hat{\zeta}\hat{\omega}){\bf f}(x)=0,\;\; 
(x=x^0,\;{\bf x})$.   
Using it in (\ref{T}), one can verify
that both components $f_\pm (x)$ of the state vector obey
one and the same equation
\begin{equation}
(\imath\partial_\mu\Gamma^\mu_s-sm)f_\zeta(x)=0\;,\;\;\zeta =\pm 1\;,
\end{equation}
which is the 2+1 Dirac equation for a particle of spin $s/2$, whereas 
$f_\pm(x)$ can be interpreted as positive and negative frequency solutions 
of the equation. Substituting the realization (\ref{fcc1}) into the
expression for the generators of the Lorentz transformations, we get 
$\hat{M}_{\mu\nu}=-\imath(x_\mu\partial_\nu-x_\nu\partial_\mu)-\frac{\imath}{4}
\hbox{diag}\left([\Gamma_{s\mu},\Gamma_{s\nu}]\;,\;\;
[\Gamma_{s\mu},\Gamma_{s\nu}]\right)$, 
which have the standard form for both components $f_\zeta (x)$. Thus, a natural
interpretation of the components $f_\zeta(x)$ is the following: $f_+(x)$ is
the wave function of a particle with spin $s/2$ and $f^*_-(x)$ is the wave
function of an antiparticle with spin $s/2$. Such an interpretation can be
confirmed if we switch on an external electromagnetic field. In this case the
coupling constants with the external field in the equations for $f_\zeta(x)$
are $\zeta g$, i.e. have different sign for particle and antiparticle.

It is interesting that the model proposed can be
derived in course of a dimensional reduction from the model  
for the Weyl particle in 3+1 dimensions, constructed in \cite{GGT2}.  
As is known, the method of dimensional reduction appears to be
often useful to construct models (actions) in low dimensions
using some appropriate models in higher dimensions \cite{DNP}. 
In the gauge $\psi^0=0$ (or in any gauge linear in $\psi^\mu$) one can
see that, among the four constraints $T_\mu$ of the model \cite{GGT2} 
only one is independent. Thus, in fact, one can use only one component 
of $\kappa^\mu$ and all others put to be zero.  In 3+1 dimensions 
this violates the explicit Lorentz invariance on the classical 
level. However in 2+1 dimensions it does not. So, if we make a
dimensional reduction 3+1 $\rightarrow$  2+1 in the Hamiltonian and constraints
of the model\cite{GGT2}, putting also $\pi_3=m$, $\kappa^3\equiv\kappa$, 
whereas $\kappa^0=\kappa^1=\kappa^2=0$, then as a result of such a procedure 
we just obtain the expression (\ref{ham}) (at $A=0$) for the
Hamiltonian of the massive Dirac particle
in 2+1 dimensions and all the constraints of the latter model.  In the 
presence of an electromagnetic field one has also to put $A_3=0$,
$\partial_3A_\mu=0$ to get the same result.

Two authors, D. M. Gitman and A. E. Gon\c calves, thank
Brazilian foundations FAPESP, CNPq 
and CAPES for support and I.V. Tyutin thank 
the European Community Commission which is supporting him in part 
under the contract INTAS-94-2317.

\end{document}